\documentclass[a4paper]{jpconf}
\pdfoutput=1
\usepackage{amsmath}
\usepackage{booktabs}
\usepackage{multirow}
\usepackage{prettyref}
\usepackage{xcolor}
\usepackage[absolute]{textpos}

\input{all.tikzdefs}

\tikzstyle{blank}=[fill=white, shape=circle, draw=white, inner sep=0.8pt]
\tikzstyle{dot}=[fill=black, shape=circle, draw=black, inner sep=0.8pt]
\tikzstyle{fat}=[fill=white, shape=circle, draw={rgb,255: red,176; green,36; blue,39}, dashed, line width=1pt, inner sep=0.8pt]

\tikzstyle{edge}=[-, draw=PropagatorColor, line width=1.2pt, line cap=rect, preaction={{draw=white,line width=2.5pt}}]
\tikzstyle{incoming edge}=[line width=1pt, line cap=rect, draw={rgb,255: red,102; green,102; blue,102}, {|-}]
\tikzstyle{outgoing edge}=[line width=1pt, line cap=rect, draw={rgb,255: red,102; green,102; blue,102}, ->]
\tikzstyle{external edge}=[line width=1pt, line cap=rect, draw={rgb,255: red,102; green,102; blue,102}, -]
\tikzstyle{top}=[-, draw=TopPropagatorColor, line width=1.9pt, line cap=rect, preaction={{draw=white,line width=2.5pt}}]
\tikzstyle{edge dot1}=[-, postaction=decorate, decoration={markings,mark=at position .50 with {\node[style=dot]{};}}]
\tikzstyle{edge dot2}=[-, postaction=decorate, decoration={markings,mark=between positions 0.33 and 0.67 step 0.33 with {\node[style=dot]{};}}]
\tikzstyle{edge dot3}=[-, postaction=decorate, decoration={markings,mark=between positions 0.25 and 0.76 step 0.25 with {\node[style=dot]{};}}]
\tikzstyle{edge dot4}=[-, postaction=decorate, decoration={markings,mark=between positions 0.20 and 0.81 step 0.20 with {\node[style=dot]{};}}]
\tikzstyle{dot1}=[-, draw=none, postaction=decorate, decoration={markings,mark=at position .50 with {\node[style=dot]{};}}]
\tikzstyle{dot2}=[-, draw=none, postaction=decorate, decoration={markings,mark=between positions 0.33 and 0.67 step 0.33 with {\node[style=dot]{};}}]
\tikzstyle{dot3}=[-, draw=none, postaction=decorate, decoration={markings,mark=between positions 0.25 and 0.76 step 0.25 with {\node[style=dot]{};}}]
\tikzstyle{dot4}=[-, draw=none, postaction=decorate, decoration={markings,mark=between positions 0.20 and 0.81 step 0.20 with {\node[style=dot]{};}}]
\tikzstyle{incoming}=[line width=1pt, line cap=rect, draw=LegColor, {|-}]
\tikzstyle{outgoing}=[line width=1pt, line cap=rect, draw=LegColor, ->]
\tikzstyle{outgoing top}=[line width=1pt, line cap=rect, draw=TopLegColor, ->]
\tikzstyle{outgoing higgs}=[line width=1pt, line cap=rect, draw=HiggsLegColor, ->]
\tikzstyle{edge}=[-, draw={rgb,255: red,176; green,36; blue,39}, line width=1pt, preaction={{draw=white,line width=2pt}}, line cap=rect]
\tikzstyle{massive edge}=[-, draw={rgb,255: red,133; green,119; blue,181}, line width=2.0pt, preaction={{draw=white,line width=2.5pt}}, line cap=rect]
\tikzstyle{cut edge}=[-, draw={rgb,255: red,64; green,64; blue,64}, line width=0.5pt, densely dashed, line cap=rect]
\tikzstyle{xcut edge}=[-, draw={rgb,255: red,64; green,64; blue,64}, line width=0.5pt, densely dashed, line cap=rect, postaction={decorate, decoration={markings,mark=at position .50 with {\node[cross out,solid,draw=white,line width=2pt,inner sep=1.4pt,transform shape] {};}}}, postaction={decorate, decoration={markings,mark=at position .50 with {\node[cross out,solid,draw={rgb,255: red,176; green,36; blue,39},line width=1pt,inner sep=1.8pt,transform shape] {};}}}]
\tikzstyle{xcut edge 1/3}=[-, draw={rgb,255: red,64; green,64; blue,64}, line width=0.5pt, densely dashed, line cap=rect, postaction={decorate, decoration={markings,mark=at position .33 with {\node[cross out,solid,draw=white,line width=2pt,inner sep=1.4pt,transform shape] {};}}}, postaction={decorate, decoration={markings,mark=at position .33 with {\node[cross out,solid,draw={rgb,255: red,176; green,36; blue,39},line width=1pt,inner sep=1.8pt,transform shape] {};}}}]
\tikzstyle{cut}=[-, draw={rgb,255: red,61; green,171; blue,83}, line width=0.7pt, dotted, line cap=rect]
\tikzstyle{photon}=[-, draw={rgb,255: red,176; green,36; blue,39}, line width=1pt, preaction={{draw=white,line width=2pt}}, line cap=rect, decorate, decoration=snake]
\tikzstyle{gluon}=[-, draw={rgb,255: red,176; green,36; blue,39}, line width=1pt, preaction={{draw=white,line width=2pt}}, line cap=rect, decorate, decoration={coil,aspect=1.4,segment length=2.5mm}]
\tikzstyle{gluoncoil}=[-, decorate, decoration={coil,aspect=1.4,segment length=2.5mm}]
\tikzstyle{fermion}=[-, draw={rgb,255: red,176; green,36; blue,39}, line width=1pt, preaction={{draw=white,line width=2pt}}, line cap=rect, postaction=decorate, decoration={markings,mark=at position .60 with {\arrow{stealth[round]}}}]
\tikzstyle{ghost}=[-, style=fermion, line width=1pt, line cap=round, dash pattern={on 0pt off 3\pgflinewidth}]
\tikzstyle{scalar}=[-, line width=1pt, densely dashed, draw={rgb,255: red,102; green,102; blue,102}]
\tikzstyle{fermionarrow}=[-,postaction=decorate, decoration={markings,mark=at position .60 with {\arrow{stealth[round]}}}]
\tikzstyle{arrow}=[{-{Classical TikZ Rightarrow[length=2mm,width=1.5mm]}}, draw={rgb,255: red,61; green,171; blue,83}, line width=1pt, preaction={{draw=white,line width=2pt}}, line cap=rect]
\tikzstyle{brace}=[-,draw={rgb,255: red,61; green,171; blue,83}, line width=1pt, decorate, decoration={brace,amplitude=5pt}]

\usepackage[pdfusetitle]{hyperref}
\definecolor{EmeraldGreen}{HTML}{1ea78d}
\definecolor{EnglishRed}{HTML}{b02427}
\hypersetup{colorlinks=true,urlcolor=EmeraldGreen,citecolor=EmeraldGreen,linkcolor=EnglishRed}

\definecolor{ZetaLightBrown}{HTML}{bf8040}
\renewcommand{\emph}[1]{\textit{\color{ZetaLightBrown}#1}}

\let\origref\ref
\newrefformat{sec}{\hyperref[#1]{Section~\origref*{#1}}}
\newrefformat{tab}{\hyperref[#1]{Table~\origref*{#1}}}
\newrefformat{fig}{\hyperref[#1]{Figure~\origref*{#1}}}
\newrefformat{eq}{eq.\,\hyperref[#1]{(\origref*{#1})}}
\AtBeginDocument{\renewcommand*{\ref}[1]{\prettyref{#1}}}

\begin{document}

\title{Loop integral evaluation and asymptotic expansion with py\textsc{SecDec}}
\author{Vitaly Magerya}
\address{Institute for Theoretical Physics, Karlsruhe Institute of Technology,}
\address{Wolfgang-Gaede-Str. 1, Geb. 30.23, 76131 Karlsruhe}
\ead{vitalii.maheria@kit.edu}

\begin{abstract}
The evaluation of higher-loop Feynman integrals is at the core of the quest to reduce the uncertainty of theoretical predictions and match experimental data from the LHC and future colliders. py\textsc{SecDec} is a program to evaluate such integrals numerically based on the sector decomposition approach; its new release version 1.5 introduces features significantly improving its performance: automatic adaptive evaluation of weighted sums of integrals (e.g. amplitudes) and asymptotic expansion in kinematic ratios. Here we briefly review both, illustrating the expected performance benefits.
\end{abstract}

\begin{textblock*}{2cm}(15.2cm,6cm)
    \hfill\normalsize\texttt{ KA-TP-02-2022}
\end{textblock*}

\section{Introduction}

The increasing amount of data from the Large Hadron Collider
(LHC) allows the experimental groups to push the precision
boundaries of the knowledge of scattering cross-sections, and
poses an ongoing challenge to the theoreticians to make the
theoretical predictions at least as precise as the experimental
ones. Already now at least 2-loop QCD corrections are needed for
this, and future colliders will require 3-loop QCD and mixed
QCD-electroweak corrections~\cite{Freitas:2021oiq}.

The two major bottlenecks for these corrections are phase-space
integration and loop integration of the amplitudes. While at
1-loop level the required loop integrals are all known analytically,
and the phase-space integration can be done using e.g. the well
known Passarino-Veltman reduction~\cite{Passarino:1978jh}, no
such general solutions exist starting at the 2-loop level: the
existing phase-space integration methods are computationally
challenging and process-dependent, only a subset of the 2-loop
Feynman integrals of interest are known analytically (mostly in
the massless cases), and the numerical evaluation methods have
long-standing problems with both performance and precision.

py\textsc{SecDec}~\cite{Borowka:2017idc,Borowka:2018goh,Heinrich:2021dbf}
comes into this picture as a tool for numerical
evaluation of loop integrals, implementing the sector
decomposition approach~\cite{Binoth:2000ps,Heinrich:2008si}
(with \textsc{Fiesta}~\cite{Smirnov:2021rhf} being the
other major implementation of the same method). It has
a long history, and has started as a Mathematica code
\textsc{SecDec}~\cite{Carter:2010hi,Borowka:2012yc,Borowka:2015mxa}.
The current iteration is written in Python and C++, and is available
from GitHub\footnote{\url{https://github.com/gudrunhe/secdec}}.

Recently version~1.5 of py\textsc{SecDec} was released bringing
performance improvements through adaptive sampling of weighted
sums of integrals (i.e. amplitudes) and an implementation of
the expansion-by-regions method of the asymptotic expansion of
integrals in kinematic invariants to help with e.g. high-energy
regions where the numerical integration performance is known
to be a problem. A more detailed description can be found
in~\cite{Heinrich:2021dbf}; here we would like to report on the
current status of py\textsc{SecDec} performance when applied to
practical higher-loop calculations.

\section{The expected performance}

py\textsc{SecDec} comes with multiple configurable integrators.
Possibly the most traditionally known onw is the \textsc{Vegas}
integrator (provided by \textsc{Cuba}~\cite{Hahn:2004fe}); the
recommended one however is the Quasi Monte-Carlo integrator
(QMC) described in~\cite{Borowka:2018goh}. The major advantage
QMC has over the classical Monte-Carlo techniques, even the
advanced ones such as the \textsc{Vegas} integrator, is that its
precision scales as $1/N$ or better, with $N$ being the number
of integrand evaluations (``samples''), while the classical
approaches scale as $1/\sqrt{N}$. In other words, to get 10x
precision with QMC one needs to wait at most 10x as long, while
with classical Monte-Carlo one would have to wait 100x as long.

To illustrate the scaling and the generally expected performance,
let us turn to \ref{fig:performance} where integration times of
several 3-loop massive integrals are displayed. In short: it takes
seconds to minutes per integral to achieve practical precision.
Note that the precision scaling for the first and the second
integrals is even better than $1/N$: it is closer to $1/N^{1.5}$
and $1/N^3$ respectively, i.e. it takes less than 10x time to
achieve 10x precision.

\begin{figure}
    \centering
    \begin{tabular}{cccccccc}
    \toprule
    \multicolumn{2}{c}{\textsubscript{Diagram}$\backslash$\textsuperscript{Relative precision}} & $10^{-3}$ & $10^{-4}$ & $10^{-5}$ & $10^{-6}$ & $10^{-7}$ & $10^{-8}$\tabularnewline
    \midrule
    \midrule
    \multirow{2}{*}{$\raisebox{0.5ex}{\scalebox{0.5}{\begin{tikzpicture}
	\begin{pgfonlayer}{nodelayer}
		\node [style=none] (0) at (-2.5, 0) {};
		\node [style=dot] (1) at (-1.75, 0) {};
		\node [style=dot] (2) at (-0.75, 1) {};
		\node [style=dot] (3) at (-0.75, -1) {};
		\node [style=dot] (4) at (0.75, 1) {};
		\node [style=dot] (5) at (0.75, -1) {};
		\node [style=dot] (6) at (1.75, 0) {};
		\node [style=none] (7) at (2.5, 0) {};
		\node [style=none] (8) at (-0.25, 0) {$m_W$};
		\node [style=none] (9) at (-2, 0.25) {$m_Z$};
		\node [style=none] (10) at (2, 0.25) {$m_Z$};
	\end{pgfonlayer}
	\begin{pgfonlayer}{edgelayer}
		\draw [style=massive edge] (1) to (0.center);
		\draw [style=massive edge] (7.center) to (6);
		\draw [style=massive edge] (2) to (3);
		\draw [style=edge] (2) to (1);
		\draw [style=edge] (4) to (2);
		\draw [style=edge] (6) to (4);
		\draw [style=edge] (5) to (6);
		\draw [style=edge] (3) to (5);
		\draw [style=edge] (1) to (3);
		\draw [style=edge] (5) to (4);
	\end{pgfonlayer}
\end{tikzpicture}}}$} & GPU & 15s & 20s & 40s & 200s & 13m & 50m\tabularnewline
    \cmidrule{2-8} \cmidrule{3-8} \cmidrule{4-8} \cmidrule{5-8} \cmidrule{6-8} \cmidrule{7-8} \cmidrule{8-8}
     & \textcolor{gray}{CPU} & \textcolor{gray}{10s} & \textcolor{gray}{50s} & \textcolor{gray}{400s} & \textcolor{gray}{4000s} & \textcolor{gray}{180m} & \textcolor{gray}{1200m}\tabularnewline
    \midrule
    \multirow{2}{*}{$\raisebox{0.5ex}{\scalebox{0.5}{\begin{tikzpicture}
	\begin{pgfonlayer}{nodelayer}
		\node [style=none] (0) at (-2.5, 0) {};
		\node [style=dot] (1) at (-1.75, 0) {};
		\node [style=dot] (2) at (0, 1) {};
		\node [style=dot] (3) at (1.75, 0) {};
		\node [style=none] (4) at (2.5, 0) {};
		\node [style=dot] (5) at (0, -0.5) {};
		\node [style=dot] (6) at (-1, -0.5) {};
		\node [style=dot] (7) at (1, -0.5) {};
		\node [style=none] (8) at (-2, 0.25) {$m_Z$};
		\node [style=none] (9) at (2, 0.25) {$m_Z$};
		\node [style=none] (10) at (-1, 0.75) {$m_t$};
		\node [style=none] (11) at (1, 0.75) {$m_t$};
		\node [style=none] (12) at (1.5, -0.5) {$m_t$};
		\node [style=none] (13) at (-1.5, -0.5) {$m_t$};
		\node [style=none] (14) at (-0.5, -0.25) {$m_t$};
		\node [style=none] (15) at (0.5, -0.25) {$m_t$};
	\end{pgfonlayer}
	\begin{pgfonlayer}{edgelayer}
		\draw [style=massive edge] (0.center) to (1);
		\draw [style=massive edge] (3) to (4.center);
		\draw [style=massive edge] (1) to (2);
		\draw [style=massive edge] (2) to (3);
		\draw [style=massive edge] (3) to (7);
		\draw [style=massive edge] (7) to (5);
		\draw [style=massive edge] (5) to (6);
		\draw [style=massive edge] (6) to (1);
		\draw [style=edge, bend right=45, looseness=0.75] (6) to (7);
		\draw [style=edge] (5) to (2);
	\end{pgfonlayer}
\end{tikzpicture}}}$} & GPU & 18s & 19s & 30s & 20s & 1.2m & 2m\tabularnewline
    \cmidrule{2-8} \cmidrule{3-8} \cmidrule{4-8} \cmidrule{5-8} \cmidrule{6-8} \cmidrule{7-8} \cmidrule{8-8}
     & \textcolor{gray}{CPU} & \textcolor{gray}{5s} & \textcolor{gray}{14s} & \textcolor{gray}{60s} & \textcolor{gray}{50s} & \textcolor{gray}{12m} & \textcolor{gray}{16m}\tabularnewline
    \midrule
    \multirow{2}{*}{$\raisebox{0.5ex}{\scalebox{0.5}{\begin{tikzpicture}
	\begin{pgfonlayer}{nodelayer}
		\node [style=none] (0) at (-2.5, 0) {};
		\node [style=none] (4) at (2.5, 0) {};
		\node [style=dot] (5) at (-1.75, 0) {};
		\node [style=dot] (6) at (0, 1) {};
		\node [style=dot] (7) at (1.75, 0) {};
		\node [style=dot] (8) at (-1, -1) {};
		\node [style=dot] (9) at (1, -1) {};
		\node [style=none] (10) at (-2, 0.25) {$m_Z$};
		\node [style=none] (11) at (2, 0.25) {$m_Z$};
		\node [style=none] (12) at (1, 0.75) {$m_t$};
		\node [style=none] (13) at (1.75, -0.5) {$m_t$};
		\node [style=none] (14) at (0.25, -0.25) {$m_W$};
	\end{pgfonlayer}
	\begin{pgfonlayer}{edgelayer}
		\draw [style=massive edge] (0.center) to (5);
		\draw [style=massive edge] (7) to (4.center);
		\draw [style=massive edge] (6) to (7);
		\draw [style=massive edge] (7) to (9);
		\draw [style=massive edge] (9) to (6);
		\draw [style=edge] (5) to (8);
		\draw [style=edge] (8) to (9);
		\draw [style=edge] (8) to (6);
		\draw [style=edge] (5) to (6);
	\end{pgfonlayer}
\end{tikzpicture}}}$} & GPU & 6s & 11s & 12s & 30s & 3m & 24m\tabularnewline
    \cmidrule{2-8} \cmidrule{3-8} \cmidrule{4-8} \cmidrule{5-8} \cmidrule{6-8} \cmidrule{7-8} \cmidrule{8-8}
     & \textcolor{gray}{CPU} & \textcolor{gray}{5s} & \textcolor{gray}{10s} & \textcolor{gray}{50s} & \textcolor{gray}{800s} & \textcolor{gray}{60m} & \textcolor{gray}{800m}\tabularnewline
    \bottomrule
    \end{tabular}
    \caption{%
        \label{fig:performance}%
        py\textsc{SecDec} 1.5 integration time with Quasi
        Monte-Carlo (QMC) integrator for a selection of
        massive 3-loop electroweak self-energy integrals taken
        from~\cite{Dubovyk:2021lqe}. The CPU is AMD~Epyc~7302
        with 32 threads; the GPU is NVidia~A100.
    }
\end{figure}

\begin{figure}
    \centering
    \includegraphics[scale=0.45]{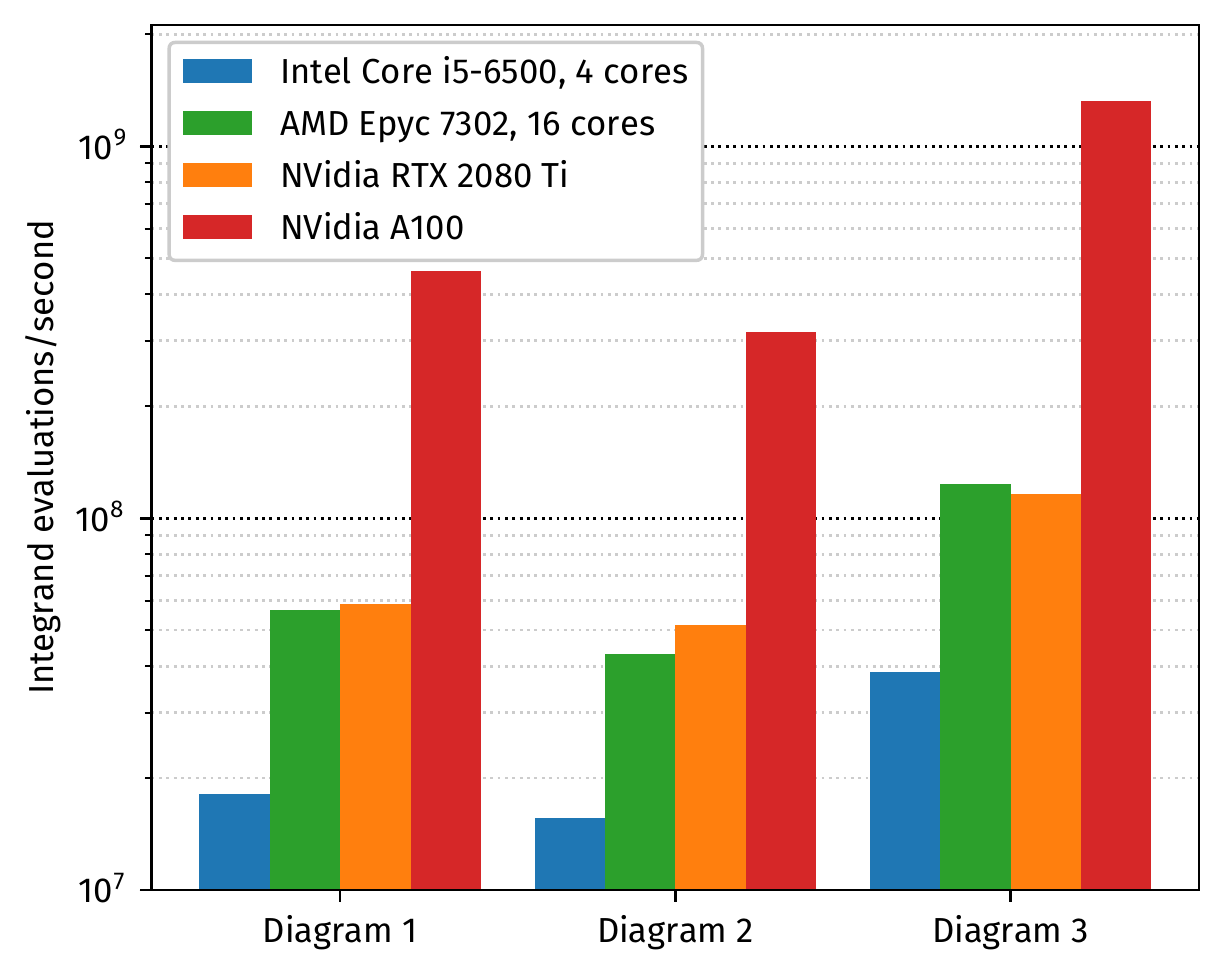}
    \caption{%
        \label{fig:evalspersecond}%
        The evaluation speed of the integrands of the diagrams
        from \ref{fig:performance} on a desktop-grade CPU (Intel
        Core i5-6500), a server-grade CPU (AMD Epyc~7302), a
        consumer-grade GPU (NVidia RTX~2080~Ti), and a server-grade
        GPU (NVidia A100).
    }
\end{figure}

Naturally, the performance depends on the employed computer
hardware. See \ref{fig:evalspersecond} for a rough comparison
between different CPU and GPU models. The takeaway is that a top
consumer-grade GPU brings the performance of a single server-grade
CPU, while a top server-grade GPU is 10x faster. For this reason we
advocate the use of server-grade GPUs for complicated integration
problems. Note that the factor limiting py\textsc{SecDec}
performance on GPUs is the use of double-precision floating
point variables---something that is not needed by the majority
of GPU users, and is not optimized for by GPU manufacturers
outside of specific server-grade GPU models like NVidia A100
or AMD MI200.\footnote{Note that py\textsc{SecDec} relies on
NVidia's CUDA libraries and therefore does not currently work
on AMD GPUs.}

\subsection{On the importance of a good integral selection}

While the integration times shown in \ref{fig:performance} are
representative of many 7- and 8-line integrals, if one needs
to evaluate a full amplitude, many more integrals would need
to be calculated, and in our experience a fraction of them
will converge much slower. For this reason it is important to
remove linearly dependent integrals by using reduction via the
integration-by-parts (IBP) relations, and it is important to
apply an effort to the selection of the master integral basis:
choosing a good basis can be the difference between having the
answer in minutes and not having it at all.

While we do not have a general recipe to predict which integrals will
converge poorly, but the rule of thumb is that one should choose
integrals with a positive power of the $U$ polynomial (in the
Feynman parametrization), and the power of $F$ should not be
lower than~$-2$. To achieve this one would sometimes need to
raise the powers of the propagators, and other times to perform
dimensional shifts (see e.g.~\cite{vonManteuffel:2014qoa}). This
does of course shift the complexity from the numerical evaluation
to the IBP reduction, but often this is an acceptable tradeoff.

\section{Adaptive sampling of amplitudes}

\begin{figure}
    \centering
    \includegraphics[viewport=0bp 8bp 575bp 272bp,clip,scale=0.45]{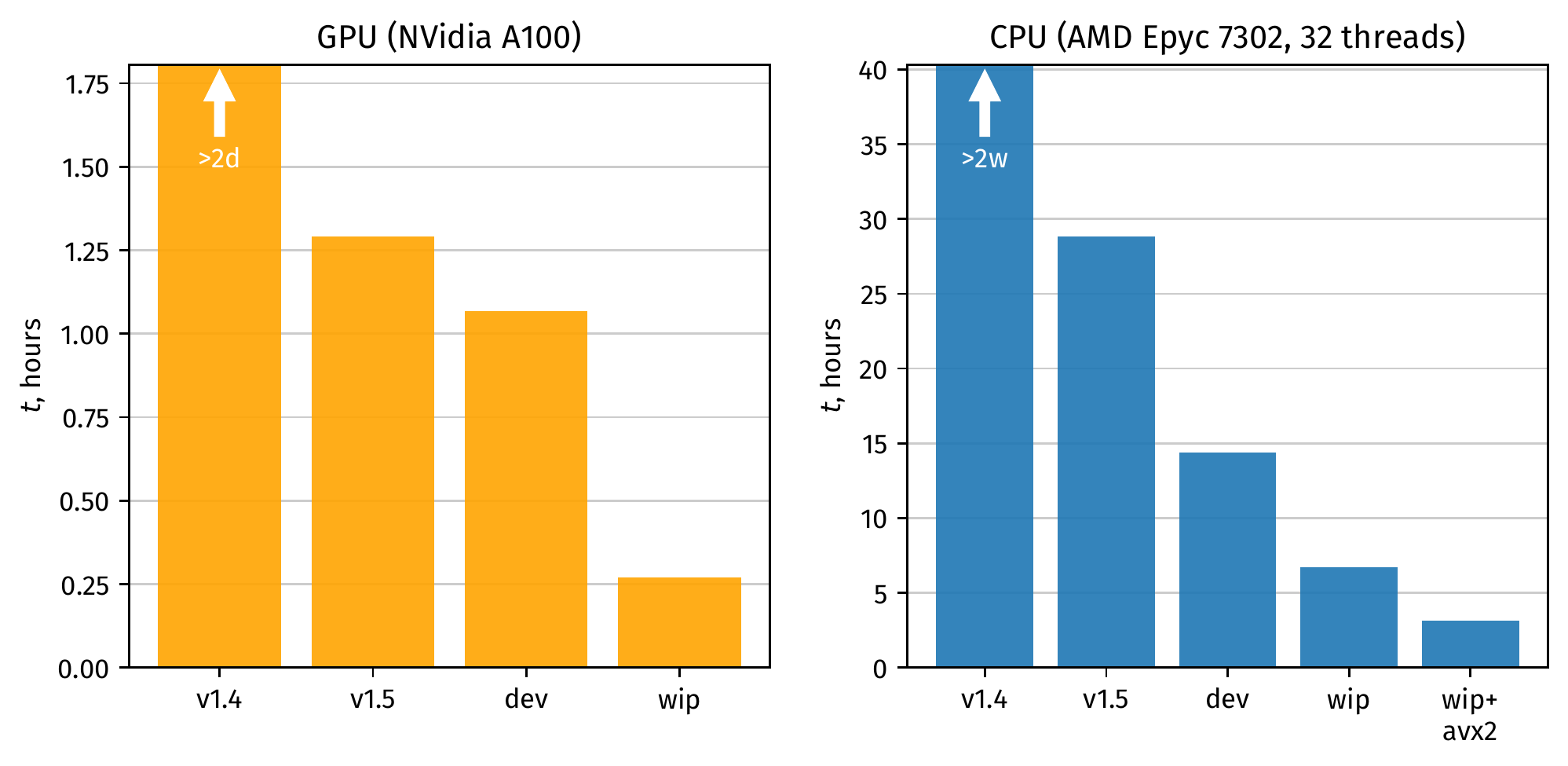}
    \caption{%
        \label{fig:perfbyversion}%
        Integration time of the first diagram from \ref{fig:performance}
        to 7 digits of precision by py\textsc{SecDec} version.
        Here \texttt{dev} stands for the current public
        development version (v1.5.3+), \texttt{wip} stands for the
        work-in-progress code to be released in the future, and
        \texttt{avx2} stands for compilation with AVX2 and FMA
        processor instruction sets allowed (the work-in-progress
        code has special provisions to use them).
    }
\end{figure}

Despite the best efforts to select a good basis of integrals,
complicated integrals can always remain. For those,
py\textsc{SecDec}~v1.5 brings a major performance improvement,
and subsequent work strives to incrementally improve on that: see
\ref{fig:perfbyversion} for a rough performance progression
of py\textsc{SecDec} versions.

The biggest contributor of the improvement between v1.4 and v1.5
is adaptive sampling of weighted sums of integrals. To see
how it works, suppose one wants to evaluate
\begin{equation}
    1\,\raisebox{0.5ex}{\scalebox{0.5}{\begin{tikzpicture}
	\begin{pgfonlayer}{nodelayer}
		\node [style=none] (0) at (-1, 0) {};
		\node [style=none] (3) at (1.25, 0) {};
		\node [style=dot] (4) at (-0.5, 0) {};
		\node [style=dot] (5) at (0.75, 0) {};
	\end{pgfonlayer}
	\begin{pgfonlayer}{edgelayer}
		\draw [style=edge, bend left=75] (5) to (4);
		\draw [style=edge, bend left=75] (4) to (5);
		\draw [style=massive edge] (5) to (3.center);
		\draw [style=edge] (4) to (5);
		\draw [style=massive edge] (0.center) to (4);
	\end{pgfonlayer}
\end{tikzpicture}
}}+10\,\raisebox{0.5ex}{\scalebox{0.5}{\begin{tikzpicture}
	\begin{pgfonlayer}{nodelayer}
		\node [style=none] (0) at (-1.25, 0) {};
		\node [style=dot] (1) at (-0.75, 0) {};
		\node [style=dot] (2) at (0, 0.5) {};
		\node [style=dot] (3) at (0, -0.5) {};
		\node [style=dot] (4) at (0.75, 0) {};
		\node [style=none] (5) at (1.25, 0) {};
	\end{pgfonlayer}
	\begin{pgfonlayer}{edgelayer}
		\draw [style=massive edge] (0.center) to (1);
		\draw [style=massive edge] (4) to (5.center);
		\draw [style=edge] (1) to (3);
		\draw [style=edge] (3) to (4);
		\draw [style=edge] (4) to (2);
		\draw [style=edge] (2) to (1);
		\draw [style=edge] (3) to (2);
	\end{pgfonlayer}
\end{tikzpicture}
}}+50\,\raisebox{0.5ex}{\scalebox{0.5}{\begin{tikzpicture}
	\begin{pgfonlayer}{nodelayer}
		\node [style=none] (0) at (-1.25, 0) {};
		\node [style=none] (3) at (1.25, 0) {};
		\node [style=dot] (4) at (-0.75, 0) {};
		\node [style=dot] (5) at (0.75, 0) {};
		\node [style=dot] (6) at (0, 0) {};
	\end{pgfonlayer}
	\begin{pgfonlayer}{edgelayer}
		\draw [style=massive edge] (5) to (3.center);
		\draw [style=massive edge] (0.center) to (4);
		\draw [style=edge, bend left=60] (4) to (6);
		\draw [style=edge, bend right=60] (4) to (6);
		\draw [style=edge, bend left=60] (6) to (5);
		\draw [style=edge, bend right=60] (6) to (5);
	\end{pgfonlayer}
\end{tikzpicture}
}}
\end{equation}
to some fixed precision. The naive approach would be to evaluate
all three integrals to this precision and then to add them.
However, because the last one has the largest coefficient and
contributes to the overall error the most, it is best to spend
more time evaluating this integral, and spend less time evaluating
the others; this way more precision can be achieved using the
same integration time. See \ref{fig:sampling} for a worked out
example.

\begin{figure}
    \centering
    \begin{tabular}{ccccc}
    \toprule
    Amplitude term & %
    \begin{tabular}{c}
    Naive\tabularnewline
    sampling\tabularnewline
    \end{tabular} & %
    \begin{tabular}{c}
    Naive\tabularnewline
    error\tabularnewline
    \end{tabular} & %
    \begin{tabular}{c}
    Better\tabularnewline
    sampling\tabularnewline
    \end{tabular} & %
    \begin{tabular}{c}
    Better\tabularnewline
    error\tabularnewline
    \end{tabular}\tabularnewline
    \midrule
    \midrule
    $1\,\raisebox{0.5ex}{\scalebox{0.5}{\input{fig/dia8master.tikz}}}$ & $10^{6}$ samples & $1\cdot10^{-6}$ & \emph{$\frac{1}{2}$}$\,\cdot\,10^{6}$ samples & $2\cdot10^{-6}$\tabularnewline
    \midrule
    $10\,\raisebox{0.5ex}{\scalebox{0.5}{\input{fig/dia1scalar.tikz}}}$ & $10^{6}$ samples & $10\cdot10^{-6}$ & \emph{$\frac{1}{2}$}$\,\cdot\,10^{6}$ samples & $20\cdot10^{-6}$\tabularnewline
    \midrule
    $50\,\raisebox{0.5ex}{\scalebox{0.5}{\input{fig/dia8master2.tikz}}}$ & $10^{6}$ samples & $50\cdot10^{-6}$ & \emph{$2$}$\,\cdot\,10^{6}$ samples & $25\cdot10^{-6}$\tabularnewline
    \midrule
    Total: & $3\cdot10^{6}$ & \emph{$51\cdot10^{-6}$} & $3\cdot10^{6}$ & \emph{$32\cdot10^{-6}$}\tabularnewline
    \bottomrule
    \end{tabular}
    \caption{%
        \label{fig:sampling}%
        Integration uncertainty of a sum of integrals for
        two different distributions of the total number of
        integrand evaluations. For illustration purposes this
        example assumes that the integration error is exactly
        $1/N_{\text{samples}}$.
    }
\end{figure}

In practice py\textsc{SecDec} will try to determine the optimal
sampling distribution not only based on the coefficients of
integrals in a sum, but also on how well they converge and how fast
can their integrand be evaluated---all determined automatically
during the integration. Note that this optimization applies to
both user-specified sums of integrals (e.g. full amplitudes) and
to single integrals too, because under the sector decomposition
method each integral is split into a sum of sectors, each sector
being a separate integral. This is why \ref{fig:perfbyversion}
shows a big improvement at v1.5 even for a single integral.

We would additionally like to note that this technology is not
new, and py\textsc{SecDec} equipped with it has previously
been successfully used in multiple 2-loop calculation such
as~\cite{Chen:2020gae,Chen:2019fla,Jones:2018hbb}.

\section{Asymptotic expansion}

Aside from some integrals being intrinsically complicated, another
source of performance problems are kinematic limits: if an
integral depends on kinematic invariants, the more extreme
their ratios get the slower such an integral converges. The reason
is that in this case the majority of the integral's value becomes
progressively concentrated in a smaller region of the integration
space, and more evaluations are needed to sample this small space
precisely.

\begin{figure}
    \centering
    \raisebox{0.5ex}{\scalebox{0.75}{\input{fig/triangle2L.tikz}}}
    \qquad
    $\vcenter{\hbox{\includegraphics[viewport=50bp 15bp 790bp 445bp,clip,scale=0.35]{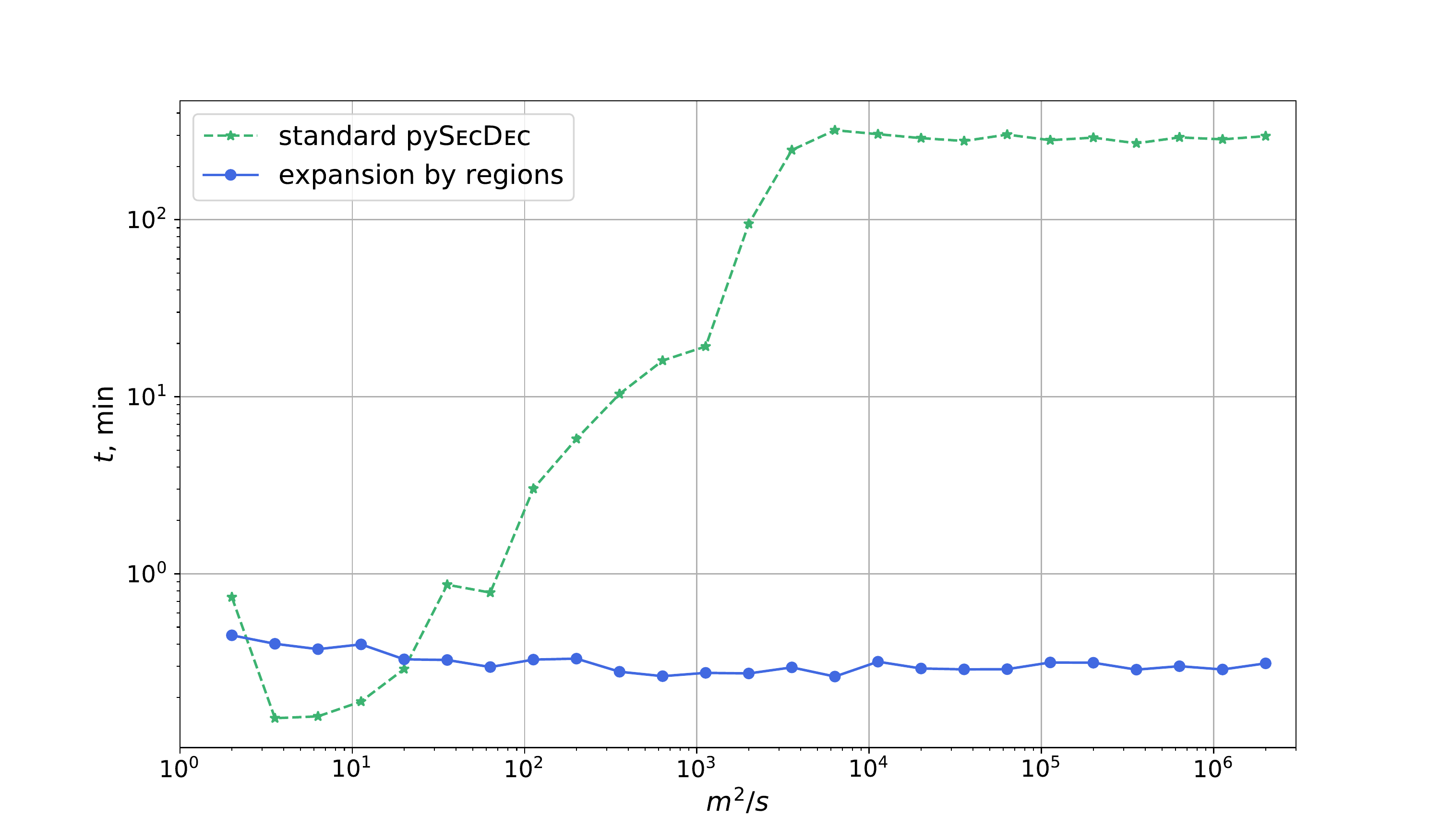}}}$
    \caption{%
        \label{fig:scan}%
        Time to evaluate the depicted integral to 3 digits of
        precision depending on the $m^2/s$ ratio. The integration
        time is capped at 5~hours, and when $m^2/s>3000$ the
        precision target can not be reached within that time.
    }
\end{figure}

As an illustration see \ref{fig:scan}: the integration time
needed to reach a given precision increases with the increase of
the $m^2/s$ ratio, eventually making it impractical to obtain
a result.

One solution to this problem is to note that since the kinematic
ratio is large, it is possible to expand the integral in the
inverse of this ratio, treating it as a smallness parameter, e.g.
\begin{equation}
    \raisebox{0.5ex}{\scalebox{0.5}{\begin{tikzpicture}
	\begin{pgfonlayer}{nodelayer}
		\node [style=dot] (0) at (-1.75, 0) {};
		\node [style=dot] (1) at (-0.75, 0.5) {};
		\node [style=dot] (2) at (0.25, 1) {};
		\node [style=dot] (4) at (-0.75, -0.5) {};
		\node [style=dot] (5) at (0.25, -1) {};
		\node [style=none] (6) at (0.75, -1) {};
		\node [style=none] (7) at (-2.25, 0) {};
		\node [style=none] (8) at (0.75, 1) {};
		\node [style=none] (9) at (-0.5, 0) {$m$};
		\node [style=none] (10) at (-2, 0.25) {$s$};
	\end{pgfonlayer}
	\begin{pgfonlayer}{edgelayer}
		\draw [style=massive edge] (7.center) to (0);
		\draw [style=external edge] (5) to (6.center);
		\draw [style=external edge] (2) to (8.center);
		\draw [style=edge] (0) to (1);
		\draw [style=edge] (1) to (2);
		\draw [style=edge] (2) to (5);
		\draw [style=edge] (5) to (4);
		\draw [style=edge] (4) to (0);
		\draw [style=massive edge] (1) to (4);
	\end{pgfonlayer}
\end{tikzpicture}
}}=\left(\cdots+\cdots\right)\left(\frac{s}{m^{2}}\right)^{-1}+\left(\cdots+\cdots+\cdots+\cdots\right)\left(\frac{s}{m^{2}}\right)^{0}+\mathcal{O}\left(\frac{s}{m^{2}}\right).
\end{equation}

The method for such expansions, ``expansion-by-regions'', has been
worked out in \cite{Beneke:1997zp,Jantzen:2011nz}: it consists
of splitting the whole integration region into subregions such
that the integration variables in each are of the specific
order in the smallness parameter, making straightforward Taylor
expansion possible; then, each of the obtained integrals can
be integrated over the whole integration region (not just its
specific subregion), with the justification that the contribution
of the overlapping subregions consist of scaleless integrals only
(and therefore is zero).

Expansion by regions was previously implemented in \textsc{Asy2.m}~\cite{Jantzen:2012mw},
and shipped as a part of \textsc{Fiesta}. As of version 1.5
py\textsc{SecDec} provides the function \texttt{loop\_regions()}
that expands a given loop integral by regions; the result of it can be directly
turned into a standard py\textsc{SecDec} integration library. The
integration time of this library is shown on \ref{fig:scan} in
comparison to the integration time of the unexpanded integral:
as expected after the expansion the integration time no longer
depends on~$m^2/s$.

\section{Conclusions}

The recently released py\textsc{SecDec} version 1.5 comes with
two major new features: automatic adaptive evaluation of the
weighted sums of integrals (e.g. amplitudes) and an implementation
of asymptotic expansion of integrals in kinematic ratios.
The first one brings a major speedup in the evaluation of single
integrals and allows using py\textsc{SecDec} to evaluate whole
amplitudes in an optimal way. The second provides an essential tool
in handling extreme kinematics (e.g. high-energy regions).

Since this release the py\textsc{SecDec} team continues incrementally
improving its performance, hoping to make it applicable to
even more challenging integrals, and to establish it as a tool to
optimally evaluate whole amplitudes. In our view sector decomposition
and py\textsc{SecDec} are the tools of the last resort: when
analytic evaluation is impossible, when differential equations are
unsolvable or unavailable, py\textsc{SecDec} will still remain
a viable option to get a result. Since the need for higher-loop
corrections is higher than ever, we believe that the time for
the last resort is nigh.

\section*{Acknowledgements}

This research was supported in part by the COST Action CA16201
(“Particleface”) of the European Union and by the Deutsche
Forschungsgemeinschaft (DFG, German Research Foundation) under
grant 396021762 (TRR 257).

\section*{References}
\bibliography{main}

\providecommand{\newblock}{}
\begin{thebibliography}{10}
\expandafter\ifx\csname url\endcsname\relax
  \def\url#1{{\tt #1}}\fi
\expandafter\ifx\csname urlprefix\endcsname\relax\def\urlprefix{URL }\fi
\providecommand{\eprint}[2][]{\href{https://arxiv.org/abs/#2}{#2}}

\bibitem{Freitas:2021oiq}
Freitas A 2021 {\em Acta Phys. Polon. B\/} {\bf 52} 929--946

\bibitem{Passarino:1978jh}
Passarino G and Veltman M~J~G 1979 {\em Nucl. Phys. B\/} {\bf 160} 151--207

\bibitem{Borowka:2017idc}
Borowka S, Heinrich G, Jahn S, Jones S~P, Kerner M, Schlenk J and Zirke T 2018
  {\em Comput. Phys. Commun.\/} {\bf 222} 313--326 (\textit{Preprint}
  \eprint{1703.09692})

\bibitem{Borowka:2018goh}
Borowka S, Heinrich G, Jahn S, Jones S~P, Kerner M and Schlenk J 2019 {\em
  Comput. Phys. Commun.\/} {\bf 240} 120--137 (\textit{Preprint}
  \eprint{1811.11720})

\bibitem{Heinrich:2021dbf}
Heinrich G, Jahn S, Jones S~P, Kerner M, Langer F, Magerya V, P\"oldaru A,
  Schlenk J and Villa E 2022 {\em Comput. Phys. Commun.\/} {\bf 273} 108267
  (\textit{Preprint} \eprint{2108.10807})

\bibitem{Binoth:2000ps}
Binoth T and Heinrich G 2000 {\em Nucl. Phys. B\/} {\bf 585} 741--759
  (\textit{Preprint} \eprint{hep-ph/0004013})

\bibitem{Heinrich:2008si}
Heinrich G 2008 {\em Int. J. Mod. Phys. A\/} {\bf 23} 1457--1486
  (\textit{Preprint} \eprint{0803.4177})

\bibitem{Smirnov:2021rhf}
Smirnov A~V, Shapurov N~D and Vysotsky L~I 2021  (\textit{Preprint}
  \eprint{2110.11660})

\bibitem{Carter:2010hi}
Carter J and Heinrich G 2011 {\em Comput. Phys. Commun.\/} {\bf 182} 1566--1581
  (\textit{Preprint} \eprint{1011.5493})

\bibitem{Borowka:2012yc}
Borowka S, Carter J and Heinrich G 2013 {\em Comput. Phys. Commun.\/} {\bf 184}
  396--408 (\textit{Preprint} \eprint{1204.4152})

\bibitem{Borowka:2015mxa}
Borowka S, Heinrich G, Jones S~P, Kerner M, Schlenk J and Zirke T 2015 {\em
  Comput. Phys. Commun.\/} {\bf 196} 470--491 (\textit{Preprint}
  \eprint{1502.06595})

\bibitem{Hahn:2004fe}
Hahn T 2005 {\em Comput. Phys. Commun.\/} {\bf 168} 78--95 (\textit{Preprint}
  \eprint{hep-ph/0404043})

\bibitem{Dubovyk:2021lqe}
Dubovyk I, Usovitsch J and Grzanka K 2021 {\em Symmetry\/} {\bf 13} 975

\bibitem{vonManteuffel:2014qoa}
von Manteuffel A, Panzer E and Schabinger R~M 2015 {\em JHEP\/} {\bf 02} 120
  (\textit{Preprint} \eprint{1411.7392})

\bibitem{Chen:2020gae}
Chen L, Heinrich G, Jones S~P, Kerner M, Klappert J and Schlenk J 2021 {\em
  JHEP\/} {\bf 03} 125 (\textit{Preprint} \eprint{2011.12325})

\bibitem{Chen:2019fla}
Chen L, Heinrich G, Jahn S, Jones S~P, Kerner M, Schlenk J and Yokoya H 2020
  {\em JHEP\/} {\bf 04} 115 (\textit{Preprint} \eprint{1911.09314})

\bibitem{Jones:2018hbb}
Jones S~P, Kerner M and Luisoni G 2018 {\em Phys. Rev. Lett.\/} {\bf 120}
  162001 (\textit{Preprint} \eprint{1802.00349})

\bibitem{Beneke:1997zp}
Beneke M and Smirnov V~A 1998 {\em Nucl. Phys. B\/} {\bf 522} 321--344
  (\textit{Preprint} \eprint{hep-ph/9711391})

\bibitem{Jantzen:2011nz}
Jantzen B 2011 {\em JHEP\/} {\bf 12} 076 (\textit{Preprint} \eprint{1111.2589})

\bibitem{Jantzen:2012mw}
Jantzen B, Smirnov A~V and Smirnov V~A 2012 {\em Eur. Phys. J. C\/} {\bf 72}
  2139 (\textit{Preprint} \eprint{1206.0546})

\end{thebibliography}
\bibliographystyle{iopart-num}
\end{document}